# Through Tissue Ultra-high-definition Video Transmission Using an Ultrasound Communication Channel

Zhengchang Kou, *Student Member, IEEE,* Andrew C. Singer, *Fellow, IEEE*, Michael L. Oelze, *Senior Member, IEEE*

*Abstract*—Wireless capsule endoscopy (WCE) has been widely adopted as complementary to traditional wired gastroendoscopy, especially for small bowel diseases which are beyond the latter's reach. However, both the video resolution and frame rates are limited in current WCE solutions due to the limited wireless data rate. The reasons behind this are that the electromagnetic (EM), radio frequency (RF) based communication scheme used by WCE has strict limits on useable bandwidth and power, and the high attenuation in the human body compared to air. Ultrasound communication could be a potential alternative solution as it has access to much higher bandwidths and transmitted power with much lower attenuation. In this paper, we propose an ultrasound communication scheme specially designed for high data rate through tissue data transmission and validate this communication scheme by successfully transmitting ultra-high-definition (UHD) video (3840*2160 pixels at 60 FPS) through 5 cm of pork belly. Over 8.3 Mbps error free payload data rate was achieved with the proposed communication scheme and our custom-built field programmable gate array (FPGA) based test platform.

## I. INTRODUCTION

THE WCE has been widely adopted as an endoscopic technique to image the small intestine, which is unreachable with traditional endoscopy, since it was first introduced in 2000[1]. However, the image resolution and frame rate of WCE are not as good as those of traditional endoscopy. Most current WCE solutions can only provide an image resolution of around 320*320 pixels and frame rates limited to several frames per second (fps) [2].

The reason behind the low image resolution and frame rate is the limited data rate (~1Mbps) that is achievable with traditional RF-based communication schemes in humans [3]. Some efforts have focused on enabling higher image resolution and frame rate WCE by using ultra-wideband (UWB) [4]. However, the relative higher attenuation suffered by UWB is the major obstacle for pursuing higher data rates and larger penetration depths for communications [5]. Furthermore, better antenna designs are required to achieve higher data rates for in-body communications using UWB [6]. An alternative solution is body-channel communications. In a recent study, a data rate of ~6 Mbps was achieved with human body communication, while the image resolution was still limited to 480*480 pixels [7].

The power limitation also sets the upper limit of the image resolution [2]. To enable higher image resolution with the limited battery capacity, low power image compressors were proposed [8],[9]. Wireless power transmission was also utilized to ease the power limit [10]. However, both power transmitter and receiver were too bulky to be used in clinical scenarios. Moreover, the potential health risk of using RF for wireless communications and power transmission through the human body is always a concern [11],[12],[13].

To overcome these challenges, ultrasound communication has been proposed to enable high data rate through tissue communications. The advantages of using ultrasound as a communication scheme for in-body communication is inherent in the physics of ultrasound as a mechanical wave. A much higher power limit of 720 $mW/cm^2$ is set for diagnostic ultrasound by the FDA. The attenuation of a 3 MHz ultrasound in tissue is more than 2 dB/cm lower than that of 500 MHz RF [14][15]. A recent study shows that up to 60 mW power can be transmitted with ultrasound over a depth of greater than 15 cm [16].

In our previous study, we achieved a data rate of over 30 Mbps with a focused ultrasound transducer [17]. One of the most significant differences between ultrasound communication and RF based communication is the much longer delay spread of ultrasound communication compared to that of RF based communication. A delay spread of over several hundred microseconds was measured through a gelatin phantom that was constructed to mimic human limbs [18]. In moving towards communications through biological tissues, a pair of 2-mm diameter sonomicrocrystals were used as a transmit and receive pair demonstrating over 4 Mbps communications through the abdomen of a rabbit [19]. To deal

This work was supported in part by the by the National Institutes of Health under grants R21EB024133, R21EB023403, R21EB030743 and R01CA251939 (Corresponding author: Michael L. Oelze.)

All the authors are with the Department of Electrical and Computer Engineering, University of Illinois at Urbana-Champaign, Urbana, IL 61801 USA (email: zkou2@illinois.edu; acsinger@illinois.edu; oelze@illinois.edu);

In addition, Z. Kou and M.L. Oelze are affiliated with the Beckman Institute for Advanced Science and Technology, University of Illinois at Urbana-Champaign, Urbana, IL 61801 USA. M. Oelze is also with the Carle Illinois College of Medicine and Department of Bioengineering, University of Illinois Urbana-Champaign, Urbana, IL 61801 USA with a much narrower coherent bandwidth and simplify the design of receiver, orthogonal frequency division multiplexing (OFDM) was explored as the modulation scheme for ultrasound in-body communication [20]. Using OFDM, real time video transmission through a rabbit abdomen was achieved with the sonomicrocrystal transmitter and an array receiver [21]. To enable high-definition video transmission, a sonomicrocrystal transmitter was used with a 64-element receive array achieving a data rate over 15 Mbps *in vivo* [22].

In this study, we proposed a new OFDM modulation structure that is designed specifically for in-body ultrasound communications and built a custom 16 channel receive verification platform with a field programmable gate array (FPGA) and ultrasound analog front end (AFE) to enable up to 25 seconds of waveform playback and record. A clip of UHD video was encoded with an H.265 encoder and modulated to an ultrasound signal, which was then transmitted with a 2-mm miniature transducer through 5 cm of pork belly and then received with a 16-channel array and recorded. The UHD video was successfully demodulated from the recorded ultrasound signal and played.

This letter is organized as follows. In Section II, the specially designed OFDM modulation scheme and signal processing chain is described. The 16-channel verification platform and test setup are explained in Section III, followed by the results in Section IV. The letter concludes with discussion of the study findings.

## II. METHODS

### A. Modulation Scheme

Based on previous studies of the ultrasound communication channel characteristics [18],[22], we propose to use block type pilots with OFDM to provide accurate channel estimation for each subcarrier because the coherent bandwidth is less than 1 KHz [22]. We also introduce comb pilots to provide phase tracking capabilities for the receiver, allowing fine frequency correction to be performed. The resource mapping of the proposed modulation is shown in Fig. 1.

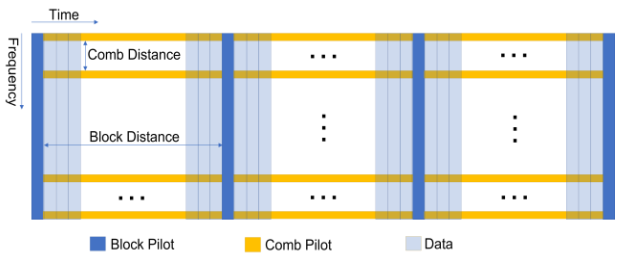

**Fig. 1. Resource mapping of the proposed OFDM modulation scheme.**

The block distance is chosen to be much smaller than the coherence time which can be calculated as $T_c \approx \frac{c}{v*f_c}$. In this case, the $T_c$ is approximately 0.2 s [2]. The number of the comb pilots is a tradeoff between pilot cost and the accuracy of fine frequency offset correction. The smaller the comb distance the more comb pilots could be used to average the residue frequency offset, which results in higher estimation accuracy.

### B. Signal Processing Chain

The UHD video (3840*2160 pixels, 60 FPS) was shot by an iPhone SE2 and was first compressed with an H.265 encoder and formatted as a binary file for channel coding, which is necessary for successfully decoding the compressed video. Low-density parity-check (LDPC) code was concatenated with cyclic redundancy check (CRC) code to form the channel coding. Then, the interleaved bits were fed into a quadrature amplitude modulation (QAM) modulator. After symbol interleaving, the complex symbols were modulated with OFDM according to the modulation scheme introduced in part II.A. A digital up converter (DUC) moved the complex baseband signal to a passband signal and a digital to analog converter (DAC) directly converted the digital passband signal to analog voltage, which was then output to the miniature transducer (sonomicrocrystal, Sonometrics, London, Ontario). The transmitter signal processing chain described above is shown in Fig. 2.

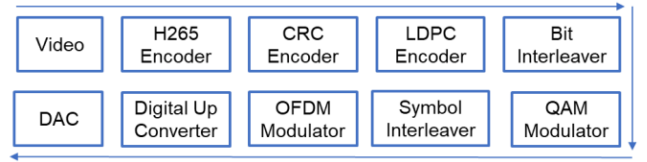

**Fig. 2. Signal processing chain of transmitter.**

On the receiver side, 16 channels of an analog front end (AFE) were used to simultaneously sample and digitize the analog signal received from 16 elements from a linear transducer array (see Fig. 3). Then, 16 identical processing modules were implemented before maximum ratio combining (MRC) the signals from each channel. In each processing module, the received digital passband signal was first down converted to a complex baseband signal by the digital down converter (DDC). Next, the cyclic prefix (CP) based coarse carrier frequency offset and symbol timing offset correction [23] were performed. Then, a fine carrier frequency offset (CFO) and symbol timing offset (STO) correction was performed after OFDM demodulation.

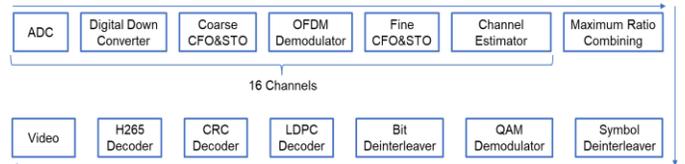

**Fig. 3. Signal processing chain of receiver.**

The fine carrier frequency offset was estimated using the comb pilots after OFDM demodulation. As the comb pilots are continuous in time, the slope of the phase of the comb pilots with respect to time can be estimated by a simple linear regression. This slope was the estimated residue carrier frequency offset.

The fine symbol timing offset was estimated according to the block pilots. According to the shift theorem of the DFT, a delay in time domain corresponds to a linear phase term in the frequency domain. A simple linear regression was used to estimate the slope of the phase of the block pilots with respect to frequency. This slope was the estimated residue time offset.

Next, a least squares channel estimation was performed on each subcarrier in the block pilot symbol followed by a linear interpolation on each data symbol. After the channel estimation, MRC was implemented as described in [22].

The equalized and combined symbols were fed into a symbol deinterleaver and QAM demodulator followed by a bit deinterleaver. Then, the LDPC decoder and CRC decoder were used to correct the errors in the binary data. Finally, error corrected data were decoded by H.265 decoder to output the UHD video.

## III. EXPERIMENT SETUP

The transmitter was based on the Redpitaya STEM 122-16, which provides a dual channel, 14 bits, 125 MHz sampling rate DAC that was directly connected to a Xilinx Zynq 7020 system-on-chip (SoC). According to the process described in part II.B., a Matlab 2022a generated complex baseband waveform was stored into a microSD card and read into the DDR3 memory (512 MB), which was directly connected to the processing system (PS) side of SoC. In the programmable logic (PL) side of SoC, a DUC was implemented to up-convert the complex baseband signal transferred from the PS side, through Advanced eXtensible Interface (AXI) direct memory access (DMA), to the real passband signal. The real passband signal was then converted to analog voltage by the DAC. The block diagram is shown in Fig. 4.

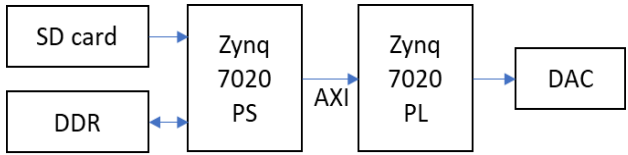

Fig. 4. Block diagram of the transmitter's hardware design.

On the receiver side, we implemented a 16-channel receiver that could record up to 25 seconds of complex baseband data on all the channels. First, a custom designed break-out board was connected to the probe to adapt the probe connector to SMA connectors. The 16 SMA cables connected the break-out board to the AFE board (TI AFE58JD48EVM). The AFE integrated both a low noise amplifier (LNA) and a programmable gain amplifier (PGA), which could provide up to 48 dB gain in total. The AFE had 16 channels with 16 bits and 125 MHz sampling rate analog to digital converters (ADCs) and digital demodulators, which could down convert the passband signal to complex baseband signal. Then, the complex baseband signal was transmitted to the PL side of Xilinx ZU7EV SoC (Xilinx ZCU106 development board) through the JESD204B high speed serial interface. A decimator was implemented inside the SoC to further reduce the data rate to prolong the recording time. The decimated data were then buffered into the DDR4 memory (4GB) connected to the SoC's PS side via AXI DMA. When the recording was finished the data on the DDR memory was written to the SD card. The block diagram is shown in Fig. 5. Then the recorded data was loaded to Matlab for the processing described in part II.B. A picture of the hardware setup is shown in Fig. 6. The test parameters used in this study are listed in Table I.

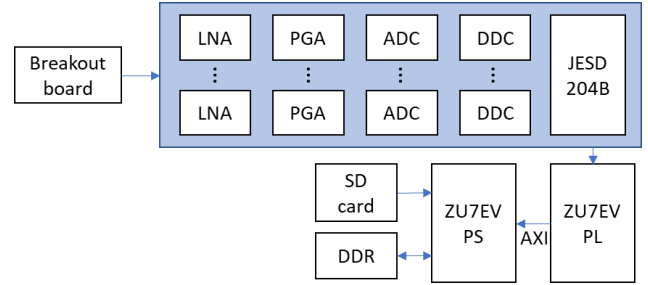

Fig. 5. Block diagram of the receiver's hardware design.

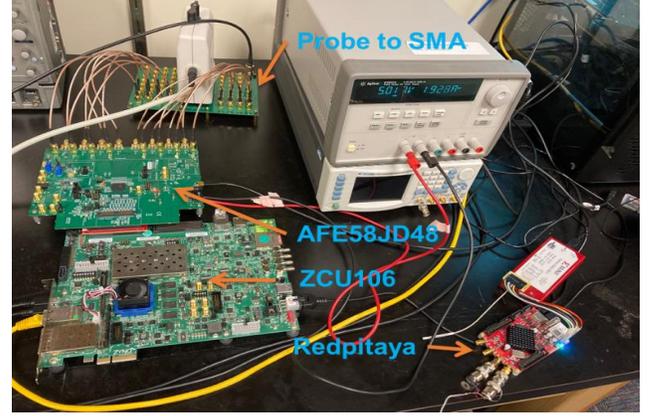

Fig. 6. Transmitter and receiver hardware setup.

**TABLE I - TEST PARAMETERS**

| Parameter | Value |
|---|---|
| Receive channel count | 16 |
| FFT size | 4096 |
| Cyclic prefix length | 512 |
| Block distance | 16 |
| Comb distance | 32 |
| Data subcarrier | 3072 |
| Modulation order | 64/256 QAM |
| LDPC code rate | 2/3 |
| DAC sampling rate | 100 MHz |
| ADC sampling rate | 120 MHz |
| Carrier frequency | 3.75 MHz |
| Baseband sampling frequency | 2.5 MHz |
| Occupied bandwidth | 1.875 MHz |
| Subcarrier spacing | 610.35 Hz |
| Encoded video bit rate | 6797 Kbps |

The low directional, miniature transducer (Sonometrics, 2-mm microcrystal) was connected to the DAC in a shielded differential fashion to minimize the electromagnetic interference through air. The receive array (IP103, Sonic Concepts, Bothell, WA) was also well shielded to minimize the coupling between the transmitter and the receiver through the air. Both the microcrystal and the receive array were submerged under water inside a water tank filled with de-ionized water. A 5-cm thick pork belly was placed on the surface of the receive probe to mimic the human abdomen channel. An additional water standoff from the microcrystal to the pork belly of 6-7 cm was also included in the setup. A photograph of the test setup is shown in Fig. 7.

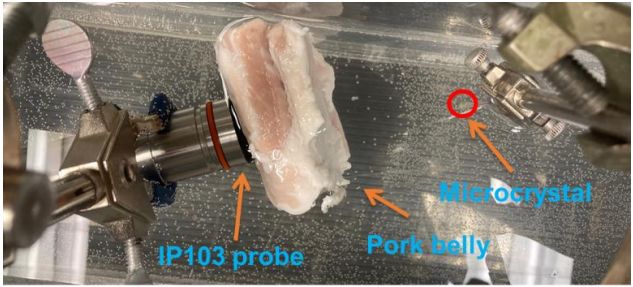

**Fig. 7. Transmit and receive transducer test setup.**

## IV. RESULTS

### A. Receiver performance

To evaluate the receiver performance, we used both 64 QAM and 256 QAM in the experiments. The constellation diagrams of 64 QAM for both single channel and 16-channel MRC receiver are shown in Fig. 8. The constellation diagrams of 256 QAM for both single channel and 16-channel receiver are shown in Fig. 9.

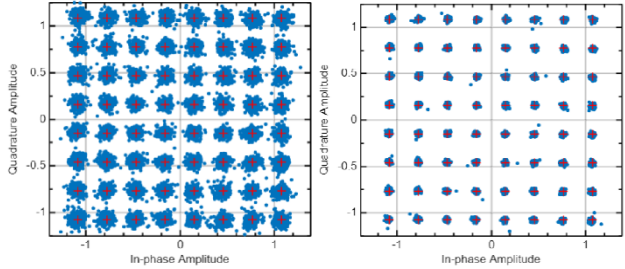

**Fig. 8. Equalized constellation diagram for 64 QAM with single channel receiver (left) and 16-channel MRC receiver (right).**

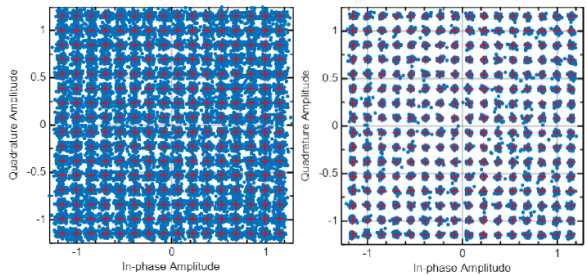

**Fig. 9. Equalized constellation diagram for 256 QAM with single channel receiver (left) and 16-channel MRC receiver (right).**

Both average and peak error vector magnitude (EVM) were measured for both 64 QAM and 256 QAM in both 16-channel MRC and single channel scenarios. The results are listed in Table II.

### B. Bit error rate performance

Bit error rate (BER) performance was evaluated by comparing the demodulated and error corrected bits with the original coded bits and encoded video bits. The BER performance for both 64 QAM and 256 QAM in both 16-channel MRC and single channel situation are shown in Table III. When there is no error within the record length, the BER is written as less than 1e-8.

**TABLE II - RECEIVER PERFORMANCE RESULTS**

|         | Average EVM (dB) |        | Peak EVM (dB) |        |
|---------|------------------|--------|---------------|--------|
|         | MRC              | Single | MRC           | Single |
| 64 QAM  | -35.29           | -27.64 | -17.13        | -8.31  |
| 256 QAM | -35.03           | -29.84 | -15.0         | -8.71  |

**TABLE III - BER PERFORMANCE RESULTS**

|         |      | Data Rate | BER    |        |
|---------|------|-----------|--------|--------|
|         | LDPC | (Mbps)    | MRC    | Single |
| 64 QAM  | Y    | 6.27      | <1e-8  | <1e-8  |
|         | N    | 9.41      | <1e-8  | 6.5e-5 |
| 256 QAM | Y    | 8.37      | <1e-8  | <1e-8  |
|         | N    | 12.55     | 3e-4   | 1e-3   |

### C. Received video

The error corrected bits were saved as a video file inside Matlab. As there is no error within the record length, the saved video file was the same as the original video that was fed into the transmitter. One frame of the video is shown in Fig. 10 (movie provided in Supplemental Video 1).

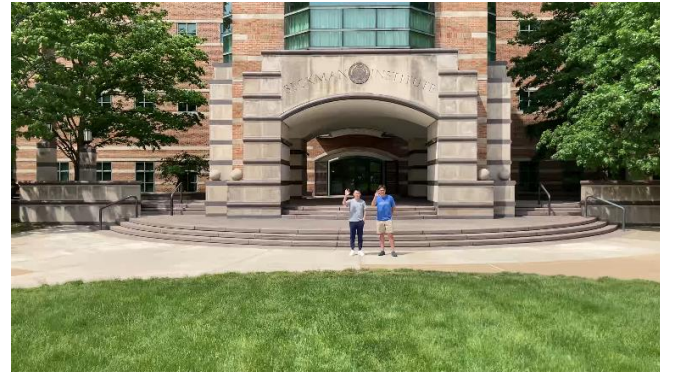

**Fig. 10. A Sample video frame saved from received video.**

## IV. DISCUSSION

In this study, we proposed a specially designed OFDM modulation scheme for ultrasound in-body communications and verified it with our custom-built FPGA-based platform. Based on the results, the performance illustrates that transmission of UHD video through an abdomen was feasible. Over 8.3 Mbps error free payload data rate was achieved with a 16-channel receiver and single channel transmitter using a miniature transmit transducer that can fit into a small capsule.

The use of comb pilots and fine CTO and SFO correction improved the receiver performance. In this study, only 16 receive channels were used, which could greatly reduce the total cost and minimize the form factor compared to previous studies using 64 channels [21],[22] and it is crucial for future portable receiver designs. Better EVM performance was achieved with the 16-channel receiver used in this study compared to the results from the 64-channel receiver used in the previous work [22].

The customized 16-channel platform also allowed the transmission of UHD video that could not be achieved in the previous study using the Verasonics ultrasound research platform (Verasonics Vantage 128). The Verasonics Vantage provides 64 MB memory for each channel, while our customized platform provides 256 MB memory for each channel. The Verasonics Vantage could only store real passband signal, which has a much higher data rate (16 bits at 62.5 MHz sampling rate) compared to the complex baseband signal (16 bits complex at 2.5 MHz sampling rate) recorded by the platform build for the current study. Up to 25 seconds of ultrasound signal could be recorded for all 16-channels with our customized platform while only 0.5 seconds could be recorded with the Verasonics Vantage 128.

The BER performance suggests that this approach and platform have a lot of redundancies for an even tougher channel scenario or higher modulation order such as 1024 QAM and higher LDPC code rate. For example, from Table III, we can observe that there was no error within the record length without LDPC coding for 64 QAM with 16-channel MRC receiver and there was no error within the record length with LDPC coding for 256 QAM with a single channel receiver. Therefore, in future work higher modulation orders and higher code rates could be implemented to achieve higher data rates. In addition, in the future the whole processing chain could be fully moved to a SoC to 1) enable real-time continuous video streaming from inside of the abdomen and 2) move towards WCE by integrating the transmitter in a custom designed chip to meet the power requirements, i.e., limits, which can be increased by using ultrasound based wireless power transmission.

ACKNOWLEDGMENT

Xilinx University Program (XUP) donated the Xilinx ZCU106 development board. Texas Instruments donated the AFE58JD48EVM AFE evaluation board.